\documentclass[sigconf,10pt]{acmart}

\usepackage{booktabs}
\usepackage{listings}
\usepackage{subcaption}

\acmYear{2025}\copyrightyear{2025}
\setcopyright{rightsretained}
\acmConference[OpenRIT 6G IV]{Workshop on Open Research Infrastructures and Toolkits for 6G}{September 8--11, 2025}{Coimbra, Portugal}
\acmBooktitle{Workshop on Open Research Infrastructures and Toolkits for 6G (OpenRIT 6G IV), September 8--11, 2025, Coimbra, Portugal}
\acmDOI{10.1145/3750718.3750744}
\acmISBN{979-8-4007-2108-3/25/09}

\begin{document}
\title[Parallelizing 6G-RAN Experimentation with VOTA]{VOTA: Parallelizing 6G-RAN Experimentation\\ with \underline{V}irtualized \underline{O}ver-\underline{T}he-\underline{A}ir Workloads}

\author{Chang Liu, T. D. Khoa Le, Rahul Saini, Kishor C. Joshi, George Exarchakos}
\affiliation{%
  \institution{Eindhoven University of Technology}
  \country{Eindhoven, The Netherlands}
}
\email{{c.liu3, t.d.k.le, r.saini, k.c.joshi, g.exarchakos}@tue.nl}
\renewcommand{\shortauthors}{Liu et al.}

\begin{abstract}
Testbed sharing, a practice in which different researchers concurrently develop independent use cases on top of the same testbed, is ubiquitous in wireless experimental research. Its key drawback is experimental inconvenience: one must delay experiments or tolerate compute and RF interference that harms experimental fidelity. In this paper, we propose \textbf{VOTA}, an open-source, software-only testbed scaling method that leverages real-time virtualization and frequency tuning to maximize parallel experiments while controlling interference. In a demonstration of two interference-sensitive 6G use cases -- \textit{MIMO iDFT/DFT Offloading} and \textit{O-RAN DoS Attack} -- running side-by-side on a 32-core host, we showcase VOTA capabilities: \textbf{dedicated-like} results while allowing \textbf{2.67$\times$} more sharing opportunities.
\end{abstract}


\maketitle

\section{Introduction}
\label{section:introduction}
The high cost of wireless research hardware, from software-defined radios (SDRs) to high-end compute units, necessitates sharing across experiments and lab members. In this paper, we raise and answer the following question: \textit{Are current testbed sharing approaches already efficient? If not, why, and how can we improve them?}

Following an informal survey, we identify two common testbed sharing approaches: (1) \textit{Shared radio equipment}, where each lab member has a dedicated compute unit but radio equipment is shared, so “ad hoc” wireless testbeds are assembled and disassembled each time the equipment changes hands; and (2) \textit{Shared shielded nodes}, where a large testbed is partitioned into fixed, RF-shielded nodes using Faraday cages or SMA cables, each assigned to a subset of lab members who must collaboratively manage confidential research artifacts and resolve software dependency conflicts.

In a broader literature survey, we found many testbed papers detailing advanced features for new use cases, but very few discuss testbed sharing. The most notable is Colosseum \cite{colosseum}, which extends the second approach by using Linux containers to address accessibility and dependency issues. However, Colosseum containers are deployed only after prototyping, so development-stage testing needs remain unaddressed. Moreover, like other RF-shielded setups, Colosseum relies on synthetic channel emulation rather than actual spatial diversity or real-world multipath, limiting its suitability for genuine MIMO and positioning use cases.

Observing that radio equipment (SDRs and embedded UEs) is the major bottleneck -- these devices are either idle or fully used -- we propose \textbf{VOTA} (Virtualized Over-The-Air), a testbed-sharing method that multiplexes compute and spectrum slices, each dedicated to a separate experiment, over a common pool of radio equipment to maximize utilization. As detailed in \textbf{Section \ref{section:vota}}, VOTA’s novelty is twofold: (i) it repurposes Linux containers as isolated workspaces, eliminating accessibility and dependency conflicts during development; and (ii) it repurposes SDR frequency tuning as spectrum slicing, mitigating RF interference without RF shielding, enabling genuine MIMO and positioning tests. VOTA is agnostic to testbed architecture (RF-shielded, over-the-air, stationary, or mobile) and applies broadly.

We next describe the VOTA method (\textbf{Section \ref{section:vota}}) and its application to our testbed (\textbf{Section \ref{section:testbed}}), which together enable near-dedicated parallel execution of two interference-sensitive use cases: \textit{MIMO iDFT/DFT GPU Offloading} and \textit{O-RAN DoS Attack} (\textbf{Section \ref{section:demo}}). We selected these use cases because both are compute-intensive and susceptible to multiple cross-experiment interference vectors, to which VOTA is expected to isolate.

\section{VOTA Methodology}
\label{section:vota}
\begin{figure}[H]
    \centering
    \includegraphics[width=1\linewidth]{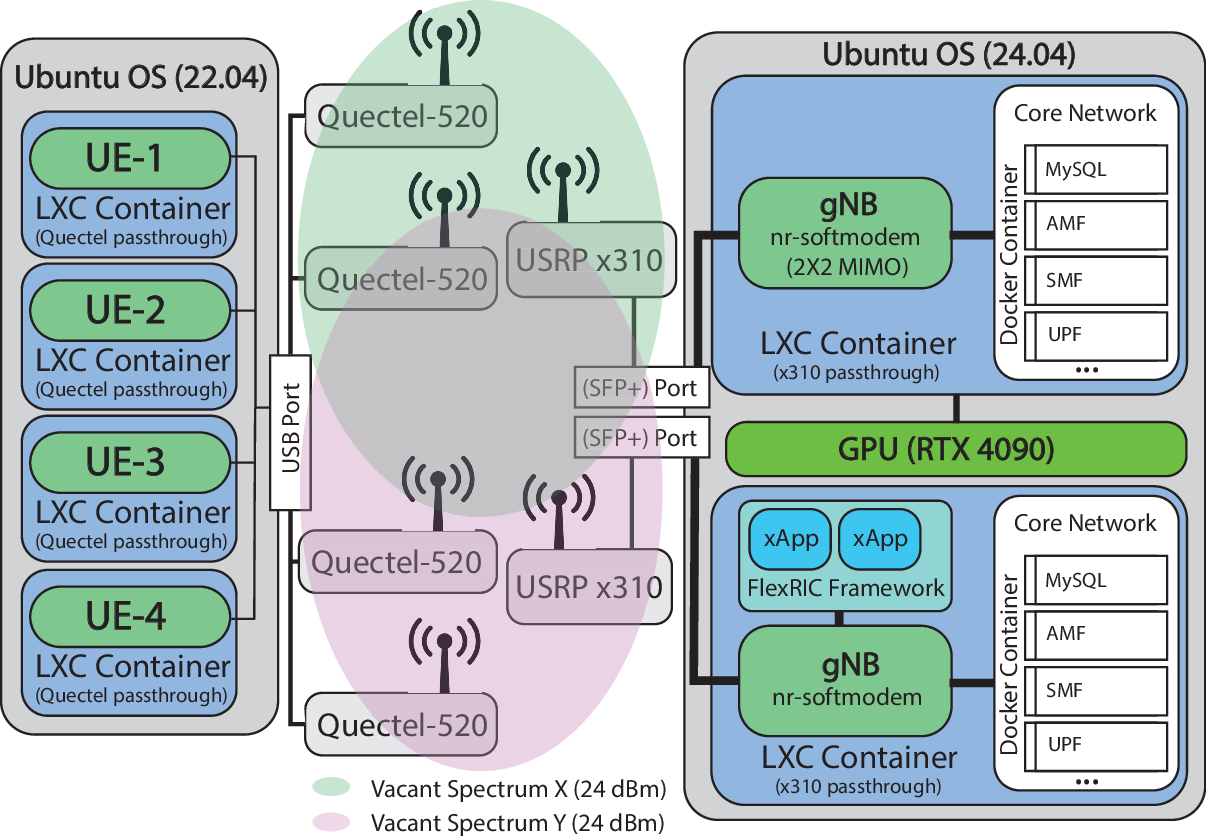}
    \caption{VOTA comprises of two core techniques: \texttt{LXC passthrough} and legal frequency tuning}
    \label{fig:architecture}
\end{figure}

Briefly stated, VOTA multiplexes compute slices (via \texttt{LXC} passthrough) and spectrum slices (by tuning operating frequencies to vacant bands within legal transmit-power limits) over a shared radio equipment pool, assigning each slice to a different experiment; see Figure~\ref{fig:architecture}. Applying VOTA to a testbed involves three steps: \emph{Isolate}, \emph{Passthrough}, and \emph{Optimize}. Before detailing these steps, we first discuss Linux Containers (LXC) -- why we chose it and how VOTA uses LXC differently.

Firstly, we pick Linux containers over virtual machines (VMs) because VMs incur extra overhead: each VM needs its own operating system kernel running on top of the host kernel. This overhead is undesirable for telecommunications workloads with strict latency requirements. In contrast, Linux containers share the host kernel and achieve isolation using kernel features such as \texttt{cgroups} \cite{cgroups} and \texttt{namespaces} \cite{namespaces}. Secondly, there are many tools to create Linux containers, notably Docker \cite{docker}, Podman \cite{Podman}, Proxmox \cite{proxmox}, and OpenStack \cite{OpenStack}. On the one hand, Docker and Podman are primarily used for deploying single applications, typically one container per app, which isn’t sufficient for development workspaces, because researchers need to run multiple applications, develop code, and persist data across restarts. On the other hand, Proxmox and OpenStack are widely used for managing distributed resources, but they are complex and require significant configuration effort. To avoid this complexity, we chose LXD \cite{LXD}, a lightweight open-source solution from Canonical for managing virtual machines and Linux containers (LXC). Within VOTA, we configure LXD with hardware passthrough, snapshots, instance duplication, live migration, and virtualized networking -- features essential for workspace isolation. Using LXD also improves multi-user security, because each container is isolated, so applications in one LXD container cannot escalate privileges into other LXD containers (Section~\ref{sub_sec:dos_atk}). Additionally, LXD includes a web UI to simplify resource management.

\textbf{Step 1a -- Compute Isolation}.
Foremost, we must isolate each user's compute workspace so that they don't interfere with each other, in both real-time and accessibility requirements. We address this by creating separate \texttt{LXD projects} \cite{lxd-project} for each user and assigning distinct CPU cores, virtual local area networks (VLANs), and storage pools to each project. Notably, the gNodeB (gNB) pins the SDR-write thread to a specific CPU core; any other tasks scheduled on that core can cause significant performance degradation or even gNB crashes. For example, when using a Universal Software Radio Peripheral (USRP), you will see repeated “LLL” (late packet) messages on the screen \cite{x310}. Therefore, it is crucial to ensure that different SDRs (and their users) are assigned to separate CPU cores. Once the projects are configured, users can create their own containers within these projects.

\textbf{Step 1b -- Spectrum Isolation}. 
Next, we apply spectrum isolation by assigning distinct unused frequency bands to each experiment, thereby minimizing interference from both commercial bands and concurrent experiments. The legality of this technique depends on local regulations, but it is as legal as conducting experiments within a Faraday cage, provided that all transmissions remain within laboratory boundaries. Using OpenAriInterface, we achieve this by configuring: (1) the downlink absolute frequency point A and the SSB position used for UE detection (\texttt{dl\_absoluteFrequencyPointA}, \texttt{absoluteFrequencySSB}), with (2) the search bands of each embedded UE (\texttt{AT+QNWPREFCFG} in the case of Quectels).

\textbf{Step 2 -- Passthrough}. Once workspace containers are created, users can ``plug-in'' any devices to their workspaces remotely. To enable this, we passthrough devices from the host to the containers. Our devices connect via USB (USRP B210, Quectel RM530N-GL, Quectel RM520N-GL) or fiber (USRP X310), and can be attached as either \texttt{usb} or \texttt{nic} devices \cite{device-passthrough}. When a device is no longer needed, it can be detached and reassigned to another user. Since we are only assigning and reassigning radio units, there is no proliferation of security control from one container to another. Additionally, since the host is a standard Linux machine, devices can be tested on the host before container assignment.

Specifically in the case of Quectel modems, there are two passthrough options. The first option is to create a VM and connect the modems by specifying the \textit{Device Number} (as all modems share the same \textit{Product ID} and \textit{Vendor ID}); however, each time the modem is physically unplugged and replugged, this \textit{Device Number} must be updated. The second option is to create a container and pass through \texttt{/dev/ttyUSB\textbf{X}}, \texttt{/dev/cdc-wdm\textbf{X}}, and \texttt{wwan\textbf{X}}, where \textbf{X} denotes the device index. To use a modem, both \texttt{/dev/ttyUSB\textbf{X}} and \texttt{wwan\textbf{X}} must be specified.

\textbf{Step 3 -- CPU Optimization}. Because a gNB requires real-time execution, the system must be configured for optimal performance. Our testbed primarily uses the USRP X310 as the base station, which demands significant processing power; thus, we aim to maximize CPU utilization for both real-time needs and to not waste resources. In addition to assigning dedicated cores to each user, we perform the following steps:
\begin{enumerate}
    \item Set all CPUs to non-idle and their frequency governors to \texttt{performance}; this locks all CPUs at their highest frequency.
    \item When running the gNB, pin its process to a specific range of cores with \texttt{taskset}, and use \texttt{nice} to grant the gNB the highest scheduling priority, minimizing interference from background processes.
    \item Search for the minimum number of cores needed to run the workload, thus increase testbed-sharing efficiency.
\end{enumerate}

\begin{table}[htb]
    \centering
    \caption{The hardware and software configuration of our experimental compute units}
    \label{tab:testbed-config}
    \begin{tabular}{@{}ll@{}}
        \toprule
        \textbf{Component}            & \textbf{Specification}                           \\ 
        \midrule
        Host Processor                & Intel Core i9‐14900K                             \\ 
        Operating System              & Ubuntu 24.04 LTS (64‐bit)                        \\ 
        Memory                        & 128 GB DDR5 RAM                                  \\ 
        Storage                       & 6 TB NVMe SSD (PCIe Gen 4)                       \\ 
        GPU                           & NVIDIA RTX 4090 (24 GB GDDR6X)                   \\ 
        Containerization              & LXC v5.21.3                                      \\ 
        Networking                    & Bridge interface                                 \\ 
        Resource Isolation            & \texttt{cgroups}, \texttt{cpusets}                \\ 
        \bottomrule
    \end{tabular}
\end{table}

\begin{figure}[htb]
    \centering
    \includegraphics[width=1\linewidth]{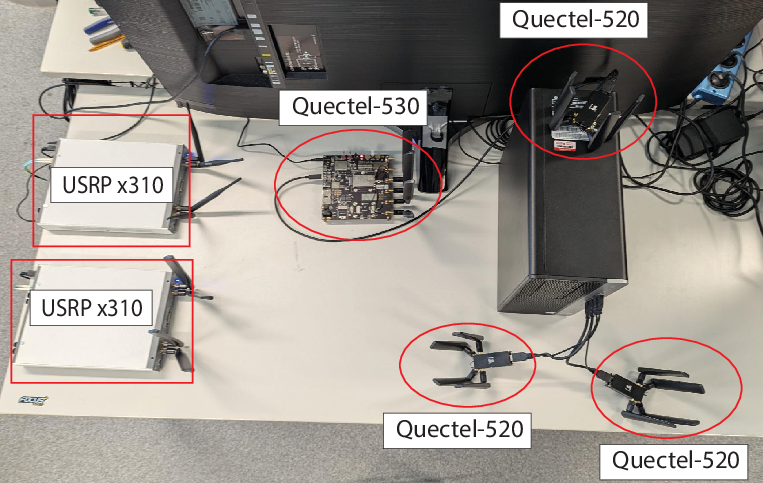}
    \caption{Our testbed is divided into a gNB cluster that connects to the same workstation (left) and a UE cluster that connects to the same PC (right)}
    \label{fig:lab}
\end{figure}

\section{Experimental Testbed}
\label{section:testbed}
Our experimental testbed utilizes a high‐performance workstation to host the gNBs, with specifications summarized in Table \ref{tab:testbed-config}. The photo of our setup are shown in Figure \ref{fig:lab}.

\section{Use Cases \& Demonstration}
\label{section:demo}
In this section, we describe our selected use cases in detail and demonstrate VOTA’s capability to run them in parallel without cross‐experiment interference. Specifically, we assign compute slice \texttt{CPU:0–11} and spectrum slice \textit{40\,MHz @\,3.32\,GHz} to Use Case 4.1, while compute slice \texttt{CPU:12–19} and spectrum slice \textit{40\,MHz @\,2.59\,GHz} are assigned to Use Case 4.2. The compute‐slicing decision was made after identifying the minimum number of cores required to run each use case, and the spectrum‐slicing decision shifts our operating frequencies to the left edge of 5G band 78 and the right edge of LTE band 41 -- both of which are not actively used in the Netherlands. Finally, we set the gNB attenuation so that OTA transmissions remain within our lab and comply with local regulations, i.e., \texttt{att\_tx = att\_rx = 8}.

\subsection{MIMO iDFT/DFT on GPU}
In 5G systems, the Inverse Discrete Fourier Transform (iDFT) and Discrete Fourier Transform (DFT) are key to Orthogonal Frequency Division Multiplexing (OFDM): iDFT at the transmitter converts frequency-domain symbols to time-domain signals, while DFT at the receiver reverses this for demodulation.

Given that iDFT/DFT operations are inherently parallelizable, offloading these transforms to a Graphics Processing Units (GPU) which has  thousands of parallel cores can yield potential performance gains. We implemented a GPU version of iDFT/DFT on top of OpenAirInterface (OAI) \cite{oai} and compared it with a highly optimized AVX version of iDFT/DFT. 

\begin{figure}[H]
    \centering
    \includegraphics[width=1\linewidth]{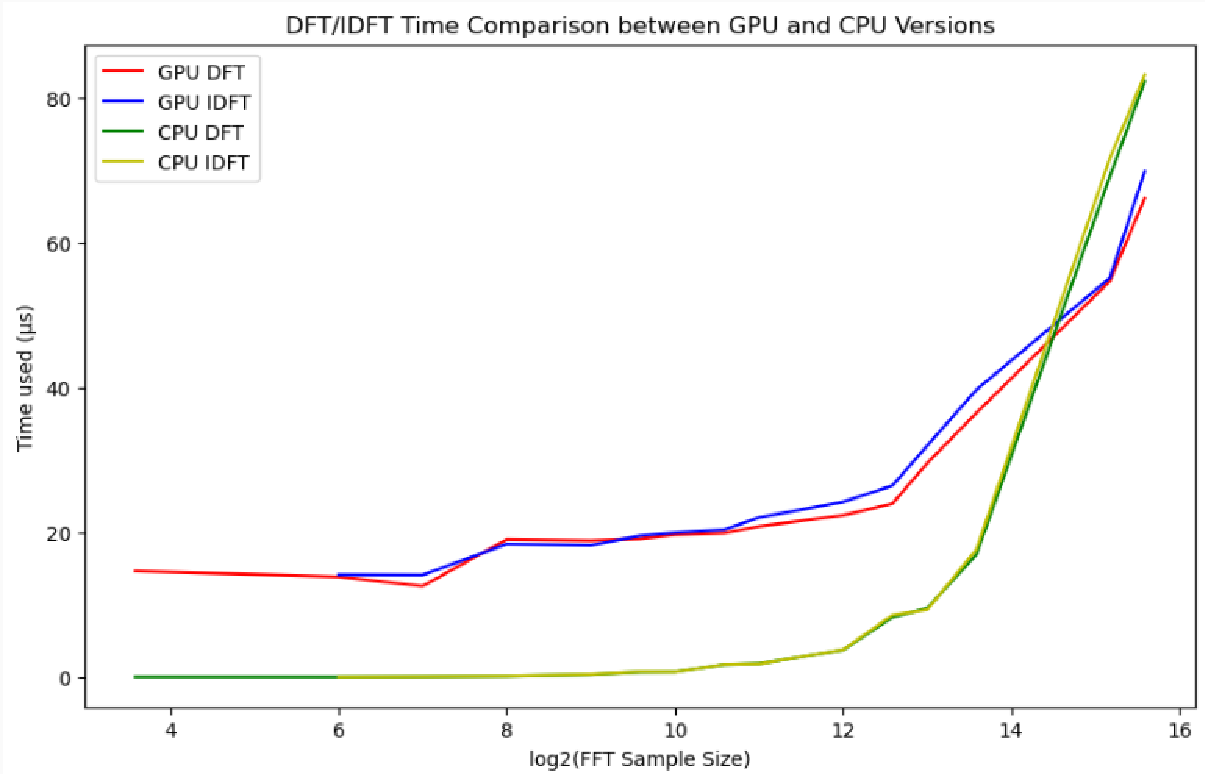}
    \caption{iDFT/DFT on CPU vs GPU}
    \label{fig:fft_cpu_gpu}
\end{figure}

\begin{figure}[H]
    \centering
    \includegraphics[width=1\linewidth]{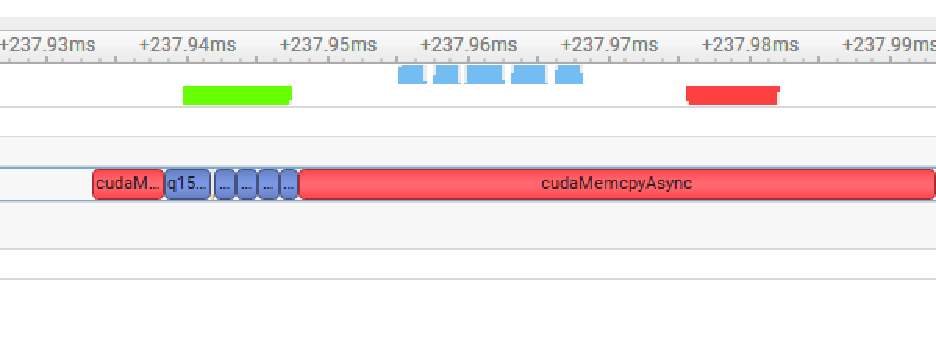}
    \caption{GPU-based iDFT/DFT implementation visualized by NVIDIA Nsight Systems\cite{nsight_system}}
    \label{fig:nsight_system}
\end{figure}

Fig. \ref{fig:fft_cpu_gpu} shows the execution time of CPU and GPU per DFT sizes \{64, 128, 256, 512, 768, 1024, 1536, 2048, 4096, 6144, 8192, 12288, 36864, 49152\}. We can see that CPU outperforms GPU when the sample size is smaller than 12288. However, per 3GPP TS38.104, there are at most $3335$ frequency ``bins'' for iDFT/DFT to process in 5G NR, dictating the maximum DFT size at $2^{\left\lceil\log _2 3335\right\rceil}=2^{12}=4096$, which means CPU should perform better in DFT for all typical 5G systems. A further investigation shown in Fig. \ref{fig:nsight_system} also proves that the bottleneck is the memory copying between RAM and VRAM, i.e. the PCIe speed. This is an interesting insight within the current research trend of GPU-based PHY layer processing.

\subsection{O-RAN DoS Attack}
\label{sub_sec:dos_atk}
This experiment employs the FlexRIC framework \cite{flexric_gitlab} to implement and integrate an xApp within the Near‐Realtime RAN Intelligent Controller (Near‐RT RIC). FlexRIC provides a hosting environment for xApps-modular, vendor‐agnostic applications deployed on the Near‐RT RIC to optimize RAN parameters. The O‐RAN architecture enables this interoperability through standardized open interfaces, fostering a competitive multi‐vendor ecosystem.

The Near‐RT RIC serves as a control entity responsible for time‐sensitive decision‐making, typically operating within a latency window of 10 ms to 1 s. Consequently, any xApp deployed on the Near‐RT RIC must satisfy these near-realtime execution constraints.

\begin{figure}[ht]
    \centering
    \includegraphics[width=\linewidth]{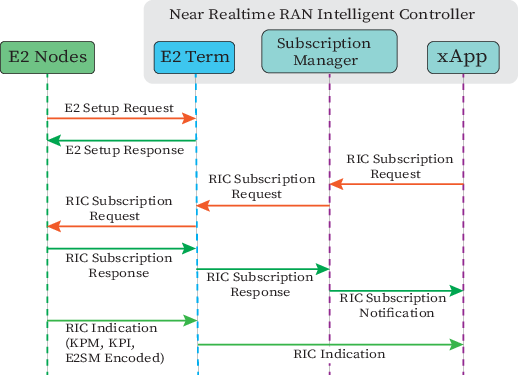}
    \caption{Typical E2AP message sequence for establishing an xApp–E2 connection}
    \label{fig:e2_flow}
\end{figure}

As illustrated in Fig.~\ref{fig:e2_flow}, the xApp initiates the interaction by sending a \textit{RIC Subscription Request} to the E2 node. Upon validation, the E2 node replies with a \textit{RIC Subscription Response}, followed by a \textit{RIC Subscription Notification} that confirms the E2 node will begin streaming relevant RAN analytics via \textit{RIC Indication} messages. This handshake is required to enable data exchange between the xApp and the E2 node.

These E2AP exchanges contain a protocol vulnerability that can be exploited to launch a Denial‐of‐Service (DoS) attack. Specifically, an xApp can repeatedly issue \textit{RIC Subscription Request} messages, overwhelming the E2 node’s processing capacity and causing it to crash \cite{DOS-ORAN}.

\begin{figure}[ht]
    \centering
    \includegraphics[width=\linewidth]{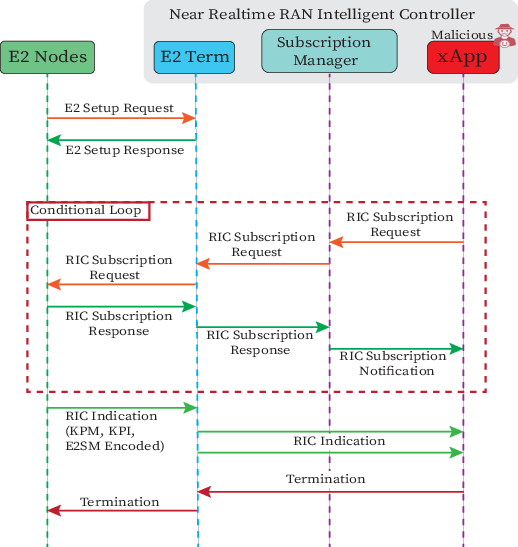}
    \caption{Protocol vulnerability exploited to achieve an O‐RAN DoS attack}
    \label{fig:atk_flow}
\end{figure}

The base station in our testbed is a USRP X310 running the \texttt{nr-softmodem} application inside an LXC container. The Core Network and FlexRIC also hosted within the same container (see Fig.~\ref{fig:architecture}). We use a monolithic gNB implementation, with that gNB instance serving as the E2 node. Because FlexRIC and the E2 node coexist in the same container, there is effectively zero network overhead for their communication. Quectel modules act as UEs to establish a connection with the experimental setup.

Our hypothesis focuses on vulnerabilities originating from xApps, which might be third‐party software solutions hosted by an operator. In this scenario, we achieve gNB service disruption by flooding the E2 node with \textit{RIC Subscription Request} messages. We measure the time required to trigger the DoS condition at the E2 node. The attack exploits the E2 node’s limited processing window (on the order of a few seconds) during which multiple xApps send \textit{RIC Subscription Request} messages. Once that window elapses and the E2 node cannot process all incoming requests, it crashes or becomes unresponsive. Despite its simplicity, this DoS method effectively impairs the gNB’s ability to perform essential control and optimization functions via the Near‐RT RIC.

\subsection{Parallel Experimentation Demo}
In this subsection, we deploy both use cases side‐by‐side and test them in parallel on our testbed. The objective is to determine whether any events of one use case -- normal or abnormal -- can affect the "ideal" testing condition of the other use case. If no cross‐experiment interference is observed, we will conclude that VOTA has successfully achieved interference‐free experimentation.

\begin{figure}[ht]
    \centering
    \includegraphics[width=\linewidth]{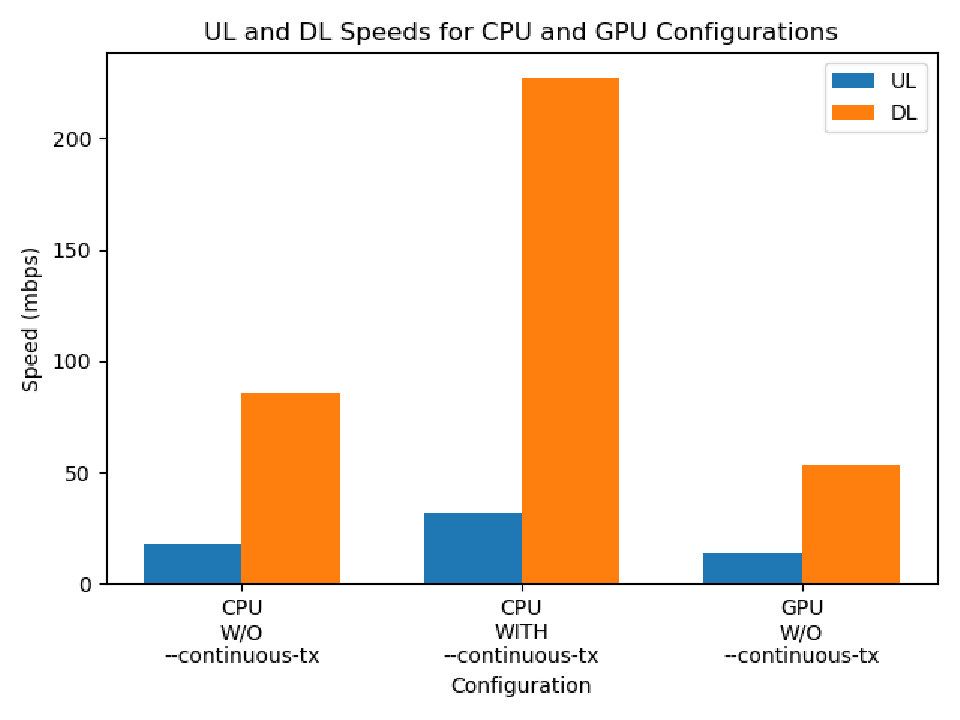}
    \caption{Over-the‐air (OTA) throughput comparison between CPU and GPU implementations.}
    \label{fig:ota_cpu_gpu}
\end{figure}

\begin{figure}[ht]
    \centering
    \includegraphics[width=\linewidth]{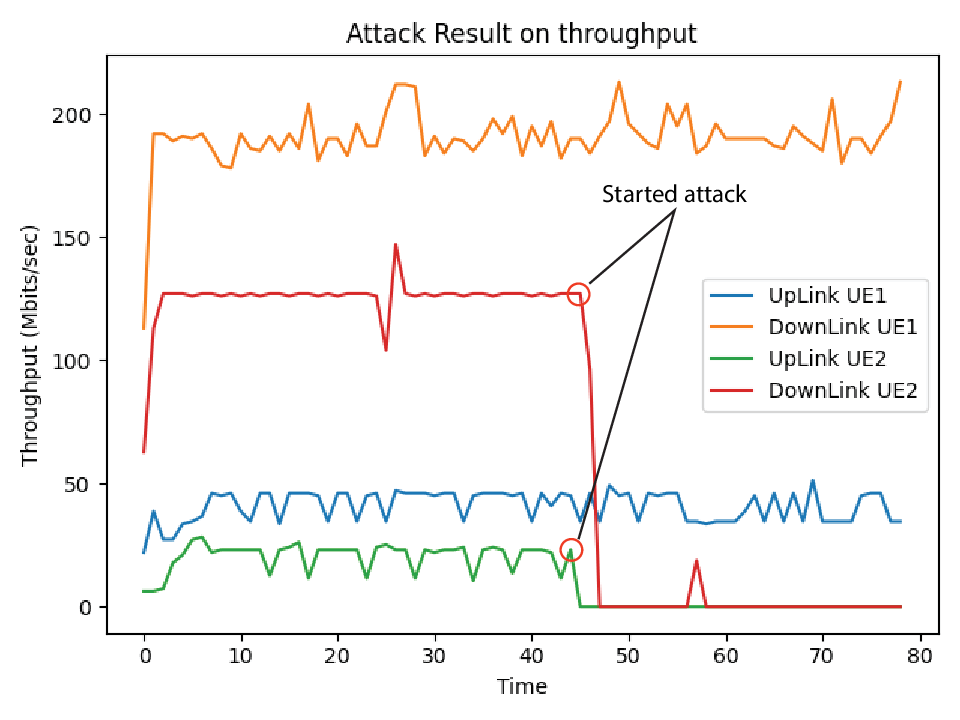}
    \caption{UE1 connected to a standard gNB using a MIMO configuration on CPU, while UE2 connected to a compromised gNB targeted by a malicious xApp. The corresponding graph shows the impact of the xApp‐initiated attack on both gNBs}
    \label{fig:atk_gnb}
\end{figure}

\textbf{CPU- vs GPU-based iDFT/DFT.} Figure~\ref{fig:ota_cpu_gpu} shows the OTA throughput results for both CPU and GPU implementations across different configurations (CPU vs GPU with/without the \texttt{--continuous-tx} option). The x‐axis labels the configuration and the y‐axis indicates the achieved system throughput. In OAI --particularly when using a USRP X310 -- enabling \texttt{--continuous-tx} is crucial to avoid power leakage from TX into RX and thus achieve high throughput. However, \texttt{--continuous-tx} requires very low processing latency, as there is no time to clear the X310’s SDR buffer between consecutive TDD frames. When attempting GPU‐based DFT/iDFT under \texttt{--continuous-tx}, we observed late‐packet (“LLL”) issues from the USRP, indicating the GPU implementation cannot meet the required latency. As shown, the CPU version outperforms the GPU version in both downlink (DL) and uplink (UL) throughput, which aligns with our findings in Section 4.1. Notably, this experiment did not experience interference from Use Case 4.2: the CPU implementation with \texttt{--continuous-tx} achieves approximately 227 Mbps -- nearly twice the throughput of a SISO link inside a Faraday cage.

\textbf{Attack on E2 Node (gNB).} Figure~\ref{fig:atk_gnb} illustrates the measured UL and DL throughput obtained using \texttt{iperf3}. This experiment evaluates the coexistence of two independent 5G systems sharing the same physical resources (CPU, memory, and GPU). We aim to determine whether an E2 Node failure in one system -- induced by an attack -- affects the performance of the other system.

To simulate an adversarial scenario, we deploy a malicious xApp that targets the gNB (acting as the E2 Node). Under normal conditions, the E2 Node maintains stable performance. Upon initiating the attack, the xApp floods the E2 Node with a large volume of RIC Subscription Request messages. Because these requests exceed the E2 Node’s processing capacity within its buffer window, the node exhausts resources and crashes, creating a denial‐of‐service (DoS) condition that disrupts both the FlexRIC platform and the connected UEs. In Figure~\ref{fig:atk_gnb}, the sharp decline in UL and DL throughput following the xApp‐triggered crash confirms the effectiveness of the attack and the resulting loss of service. However, the other gNB remains both stable and optimal.

\textbf{Efficiency Gains.} Our workstation has 32 CPU cores, and the most demanding use case -- Use Case 4.1 -- requires 12 cores. Under VOTA multiplexing, we estimate at least $\frac{32}{12} = 2.67$ testbed‐sharing opportunities. Prior to VOTA, the remaining cores could not be used safely for parallel experiments without risking interference. Thus, VOTA provides an effective increase in sharing capacity from zero to 2.67×.

\section{Conclusion and Future Work}
\label{section:future_work}
In this paper, we have shown that radio equipment -- rather than compute resources or RF interferences -- is the main bottleneck in testbed sharing. We further demonstrate that the inability to multiplex over‐the‐air (OTA) workloads on shared radio equipment is the primary factor limiting testbed utilization. Indeed, our two use cases --\textit{MIMO iDFT/DFT GPU Offloading} and \textit{O‐RAN DoS Attack} -- could not execute concurrently on the same workstation without VOTA multiplexing; sharing efficiency would be 0 instead of \textbf{2.67×}.

Our future work includes:
\begin{enumerate}
    \item Running LXC on a Linux real‐time kernel;
    \item Deploying multiple clusters for federated experiments, e.g., supporting large-scale multiple sites; and
    \item Adding O‐RUs to support O‐RAN–related experiments.
\end{enumerate}
We will release our testbed configuration and setup details as open‐source for the wider research community.

\begin{acks}
This work is funded by the Dutch National Growth Fund, project ``Future Network Services''.
\end{acks}

\bibliographystyle{unsrtnat}
\bibliography{ref}

\end{document}